\documentclass[12pt]{article}
\usepackage{graphicx}
\usepackage{cite}
\newcommand{\dissum}[2]{\displaystyle \sum_{#1}^{#2}}

\newcommand{\fnd}[2]{\frac{\textstyle #1}{\textstyle #2}}
\newcommand{\xrm}[1]{{\textstyle \mbox{\rm #1}}}
\newcommand{\bm}[1]{\mbox{\boldmath $#1$}}
\newcommand{\abs}[1]{\left| #1\right|}

\begin{document} \baselineskip .7cm
\title{\bf Nonperturbative scalar-meson resonances\protect\\
with open charm and beauty}
\author{
Eef van Beveren\\
{\normalsize\it Centro de F\'{\i}sica Te\'{o}rica}\\
{\normalsize\it Departamento de F\'{\i}sica, Universidade de Coimbra}\\
{\normalsize\it P-3004-516 Coimbra, Portugal}\\
{\small eef@teor.fis.uc.pt}\\ [.3cm]
\and
George Rupp\\
{\normalsize\it Centro de F\'{\i}sica das Interac\c{c}\~{o}es Fundamentais}\\
{\normalsize\it Instituto Superior T\'{e}cnico, Edif\'{\i}cio Ci\^{e}ncia}\\
{\normalsize\it P-1049-001 Lisboa Codex, Portugal}\\
{\small george@ajax.ist.utl.pt}\\ [.3cm]
{\small PACS number(s): 12.39.Pn, 13.75.Lb, 11.55.Bq, 14.40.Lb, 14.40.Nd}
\\[1.0cm]
{\normalsize Talk prepared for}\\
{\normalsize\it Time Asymmetric Quantum Theory: the Theory of Resonances}\\
{\normalsize July 23-26 (2003)}\\
{\normalsize Centro de F\'{\i}sica das Interac\c{c}\~{o}es Fundamentais}\\
{\normalsize Instituto Superior T\'{e}cnico, Lisboa (Portugal)}\\[0.3cm]
{\small hep-ph/0312078}
}

\maketitle
\clearpage

\begin{abstract}
We predict several new bound states and resonances for $B\pi$,
$BK$, and $BD$ $S$-wave elastic scattering,
within the same model which was recently employed to explain the
$D^{\ast}_{sJ}(2317)^{+}$ meson.
In the charm-nonstrange sector we find a (virtual ?) bound state
close to the $B\pi$ threshold, and moreover two nearby resonances of
comparable width in the range 5.9 to 6.1 GeV.
We obtain bound states $B_{s0}^{\ast}$(5570) in $b\bar{s}$
and $B_{c0}^{\ast}$(6490) in $\bar{b}c$, as well as resonances at higher
energies. No $J^{P}=0^{+}$ states are found at 5.6 GeV (in $\bar{b}u/d$),
at 5.7 GeV (in $b\bar{s}$), or at 6.1 GeV (in $\bar{b}c$),
where they are predicted from pure confinement.
\end{abstract}

\section{Introduction}
On studying the vast amount of literature
\cite{scalars,morescalars}
dealing with the phenomena observed in $S$-wave meson-meson scattering, one may
wonder whether the question \em ``What is a resonance?" \em has any answer at
all. Different theoretical considerations even come to opposite conclusions
\cite{ZPC30p615,PTPS91p284,PRD58p054012,NPA688p823,PRD65p114010}
for the same phenomenon.

One reason for the difficulty to unambiguously explain
meson-meson-scattering phenomena, may stem from the following.
In atomic physics, the perturbative approach towards electromagnetic (EM)
transitions is justified, since the EM binding forces are apparently
more decisive for the energy-level splittings than photon radiation.
We may thus predict the radiation frequencies for atomic transitions
to a good accuracy, by just determining the atomic levels of the
binding forces.
Atomic states have lifetimes which are large as compared with the internal
EM frequencies. Hence, electrons complete many revolutions about the nucleus
before the system decays. Such systems can be considered stable, like planetary
ones. We may, therefore, determine the masses of excited atoms, without making
a drastic approximation, just from the EM bound-state spectrum, which procedure
is only correct for stable systems. The wave functions for very unstable
systems have to be determined by different methods.
Most probably, they can be expressed as linear combinations of the
bound-state solutions, in which case not just one, but all excitations
contribute to the transition amplitude.

In Table~\ref{stability} we express the stability of
quark-antiquark systems by estimating the number of complete cycles of
the system before decay.
For the internal frequency of $q\bar{q}$ systems, we take the
universal frequency $\omega$ of the Nij\-me\-gen model
(see Table~\ref{parameters}).
\begin{table}[htbp]
\begin{center}
\caption{The ratios of internal frequency and width, $\omega$/$\Gamma$,
for some stable and less stable mesons.
\label{stability}}
\begin{tabular}{c|c|c}
meson & width ($\Gamma$) \cite{PRD66p010001} & periods ($\omega$/$\Gamma$)\\
\hline & \\ [-0.2cm]
$\pi^{\pm}$ & 2.53 $\times 10^{-8}$ eV & 7.5 $\times 10^{15}$\\
$\pi^{0}$ & 7.84 eV & 2.4 $\times 10^{7}$\\
$K^{\pm}$ & 5.32 $\times 10^{-8}$ eV & 3.6 $\times 10^{15}$\\
$K_{L}$ & 1.27 $\times 10^{-8}$ eV & 1.5 $\times 10^{16}$\\
$K_{S}$ & 7.37 $\times 10^{-6}$ eV & 2.6 $\times 10^{13}$\\
$\eta$(547) & 1.2 keV & 1.6 $\times 10^{5}$\\
$\omega$(782) & 8.4 MeV & 22.5\\
$\rho$(770) & 149 MeV & 1.3\\
$K^{\ast}$(892) & 51 MeV & 3.7\\
$\eta$'(958) & 0.20 MeV & 941
\end{tabular}
\end{center}
\end{table}
We read from Table~\ref{stability} that pions and Kaons are very stable,
whereas the eta mesons are reasonably stable. However, strongly decaying mesons
are very unstable.

For example, we are convinced that the structure observed at energies below
1 GeV in $P$-wave $\pi\pi$ scattering, is caused by a quark-antiquark
state, the $\rho$ meson, with a mass of about 770 MeV.
But if the $\rho$ resonance stems from a state described by a quark
confined to an antiquark, then that state is far from stable.
With a size of approximately 1 fm, its internal frequency must be some 200 MeV.
However, the lifetime of the $q\bar{q}$ state is of the same order of
magnitude, indicated by its width of 150 MeV.
Consequently, the $q\bar{q}$ system associated with the $\rho$ resonance
barely completes ONE complete revolution during its existence.
We may thus wonder whether it makes sense to refer to such a phenomenon as
a state, or even a particle, while we just concluded that it does not really
exist.

Resonances in scattering are somehow supposed to stem from an underlying
spectrum of bound states which are broken up by a weaker interaction.
Such a philosophy could in principle work very well for the {\it stable}
\/mesons $\pi$ and $K$, since their formation is associated with strong
interactions, whereas their decays are due to electro-weak processes.
Nevertheless, strong decay can have dramatic effects even for the masses
of pions and Kaons \cite{PRD27p1527}.
Once these effects are fully included in the internal dynamics of pions
and Kaons, we may indeed handle electro-weak processes perturbatively
\cite{PRD44p2803}.

In atomic systems the mechanism for binding is the same as the
mechanism for transitions, namely electromagnetism.
Nevertheless, since the fine-structure constant $\alpha$ is small,
and transitions are of higher order in $\alpha$ than the binding forces,
one may still approximate EM transitions of atoms by the Born term
in a perturbative approach.
However, the strong binding force of quarks is of comparable order
of magnitude as the strong forces that cause hadronic decay.
Consequently, in order to analyse hadronic scattering data,
one must treat both aspects of strong interactions at least on the same
footing, maybe even attributing more importance to strong decay.

The consequences of such a combined approach are, for the time being, still to
be studied in effective models, which, however primitive they are and no matter
what details need be addressed in the future, have already indicated two maior
properties of hadronic resonances:

1. Upon assuming an underlying spectrum of bare mesonic states with $q\bar{q}$
degrees of freedom only, resonances in meson-meson scattering come out at
energies which are very different from the energy levels of the underlying
confinement spectrum. This has been confirmed by he Cornell group, focusing on
heavy mesonic resonances of the charmonium and beautonium systems
\cite{Cargese75p305,PRD64p114004}, the Nij\-me\-gen group, describing, in
principle, all mesonic resonances \cite{PRD27p1527,PRD21p772}, and the Lisbon
group, in the framework of a chiral quark model \cite{PRD42p1611}. All report
mass shifts, due to coupled channels, of up to several hundred MeV, which is
comparable in magnitude to the level splittings of hadronic resonances.

2. In experiment, more resonances may be observed than just those
stemming from the assumed underlying confinement spectrum.
The Nij\-me\-gen Model predicts several mesonic resonances
\cite{ZPC30p615} and even rather stable states \cite{PRL91p012003}
which do have a different origin than the majority of mesonic resonances.
\clearpage

\section{Resonances}

Although resonances have not yet been well defined by us, we turn in this
section to the issue of their description.
Let us first discuss atomic transitions.
When a photon passes through a container filled with a gas of non-excited
atoms, then it can be absorbed once its frequency matches some allowed
transition from the ground state to an excited state of the atom's specrum.
Ideally, the absorption of photons is measured as a bell-shaped fluctuation
of the light beam's intensity, when the light frequency is continuously varied.
The reason why a Breit-Wigner (BW) form rather accurately fits the shape of
this intensity fluctuation is that the Born term of the interaction dominates
the process, and moreover, the excited state is relatively stable.
Measured over a larger frequency interval, the absorption curve
for photons can be described by a sum of BW-type contributions,
each one corresponding to a specific excited state of the atom and its
respective life time.

Were the fine-structure constant $\alpha$ not small,
then not only higher-order terms in $\alpha$ would significantly contribute,
but also would the photon noticeably couple to all higher excitations.
Hence, the form of the scattering amplitude would then certainly not
coincide with BW approximations about each peak of the absorption curve.
As a consequence, summing up BW-type amplitudes is not justified for
strong interactions. This can easily be demonstrated with a simple model.

\begin{figure}[htbp]
\begin{center}
\begin{tabular}{c}
\centerline{
\scalebox{0.45}{\includegraphics{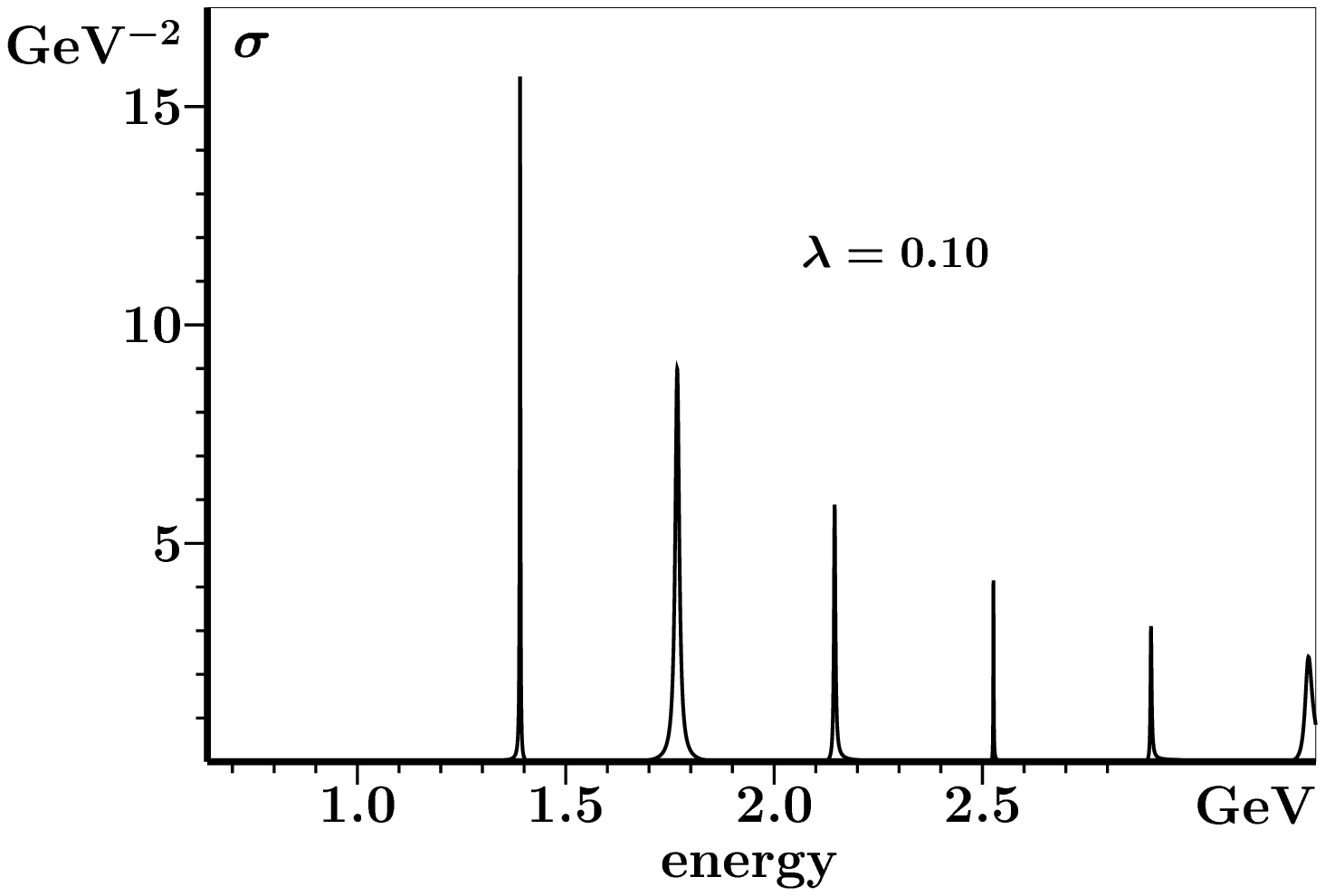}}
\scalebox{0.45}{\includegraphics{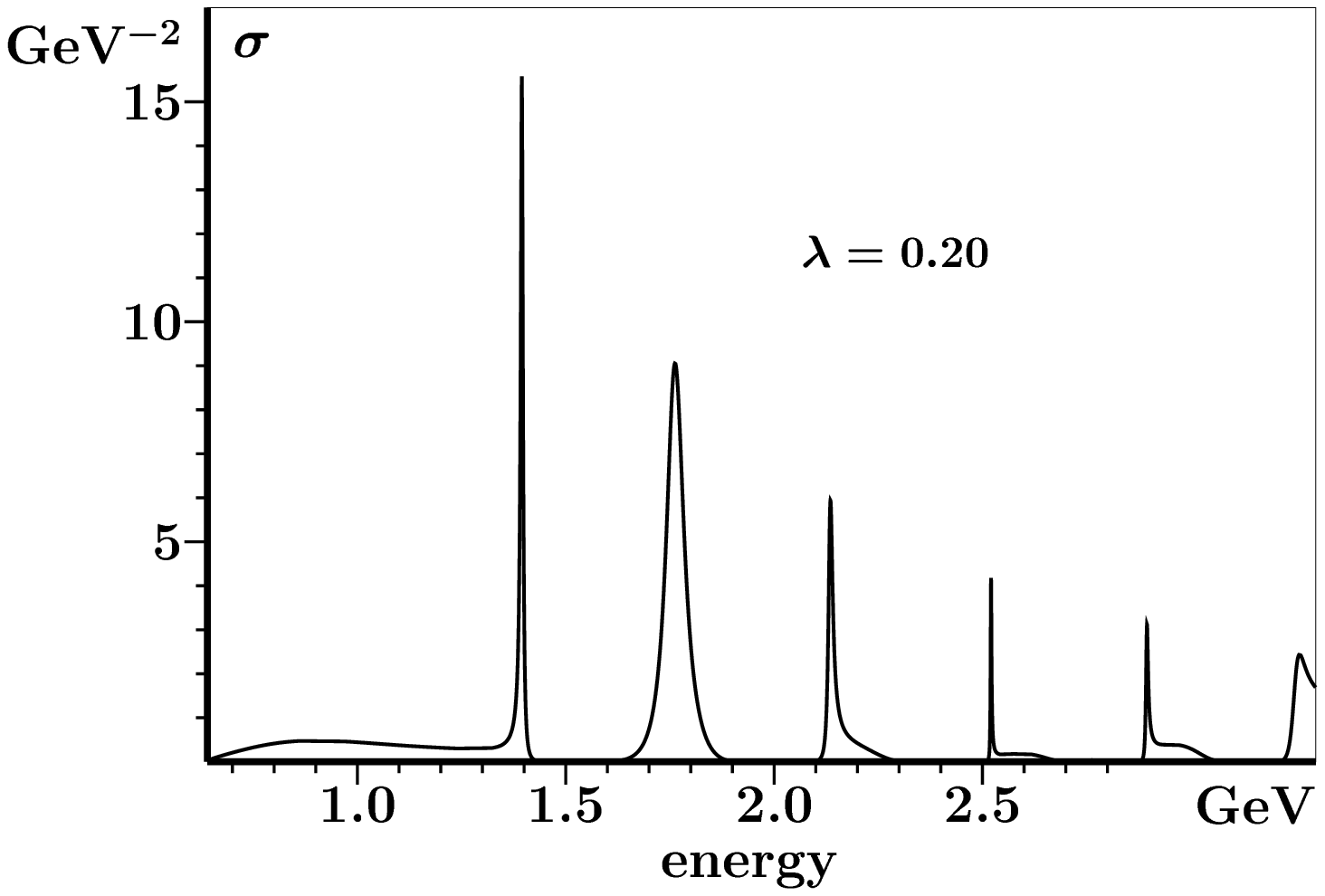}}}\\
\centerline{
\scalebox{0.45}{\includegraphics{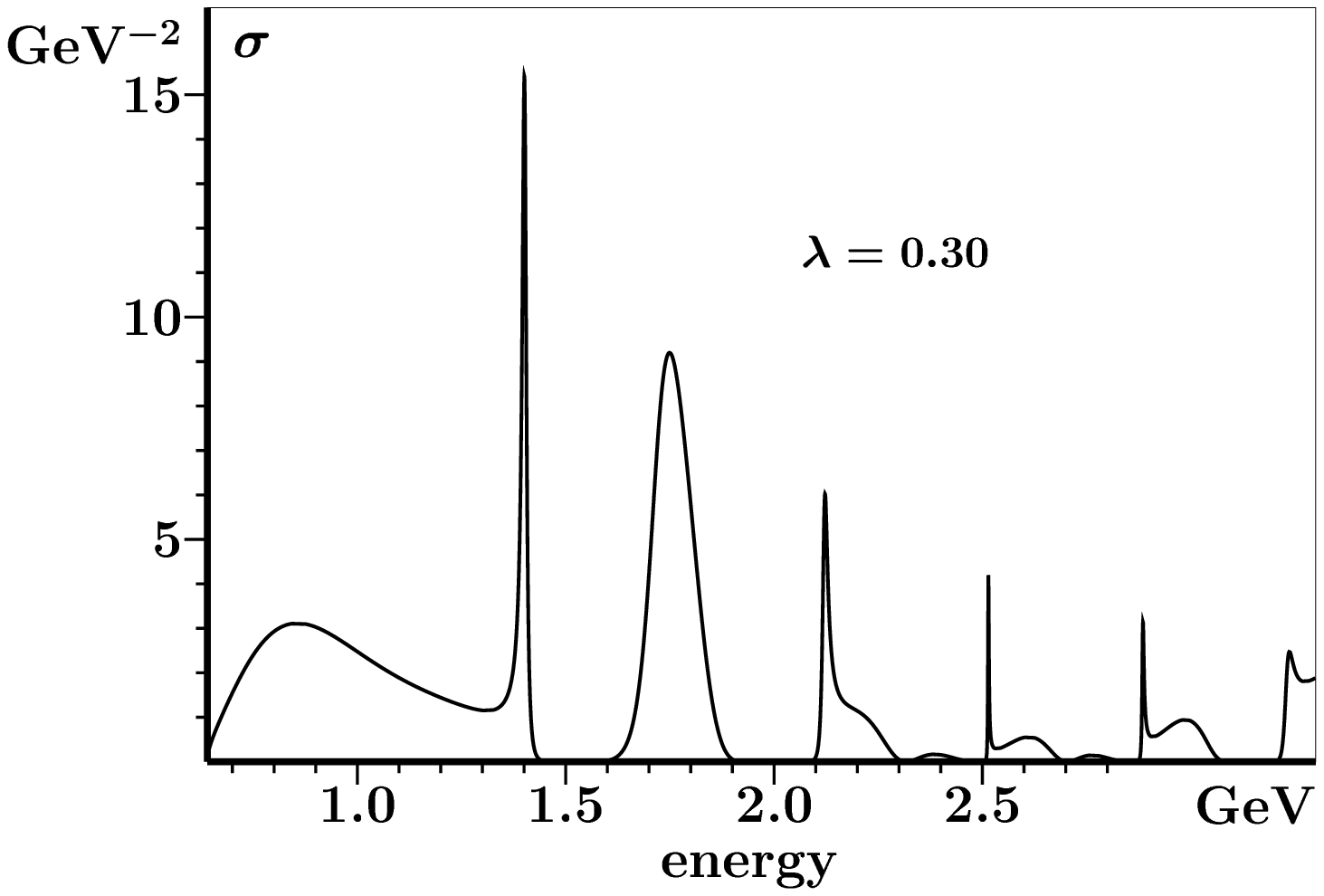}}
\scalebox{0.45}{\includegraphics{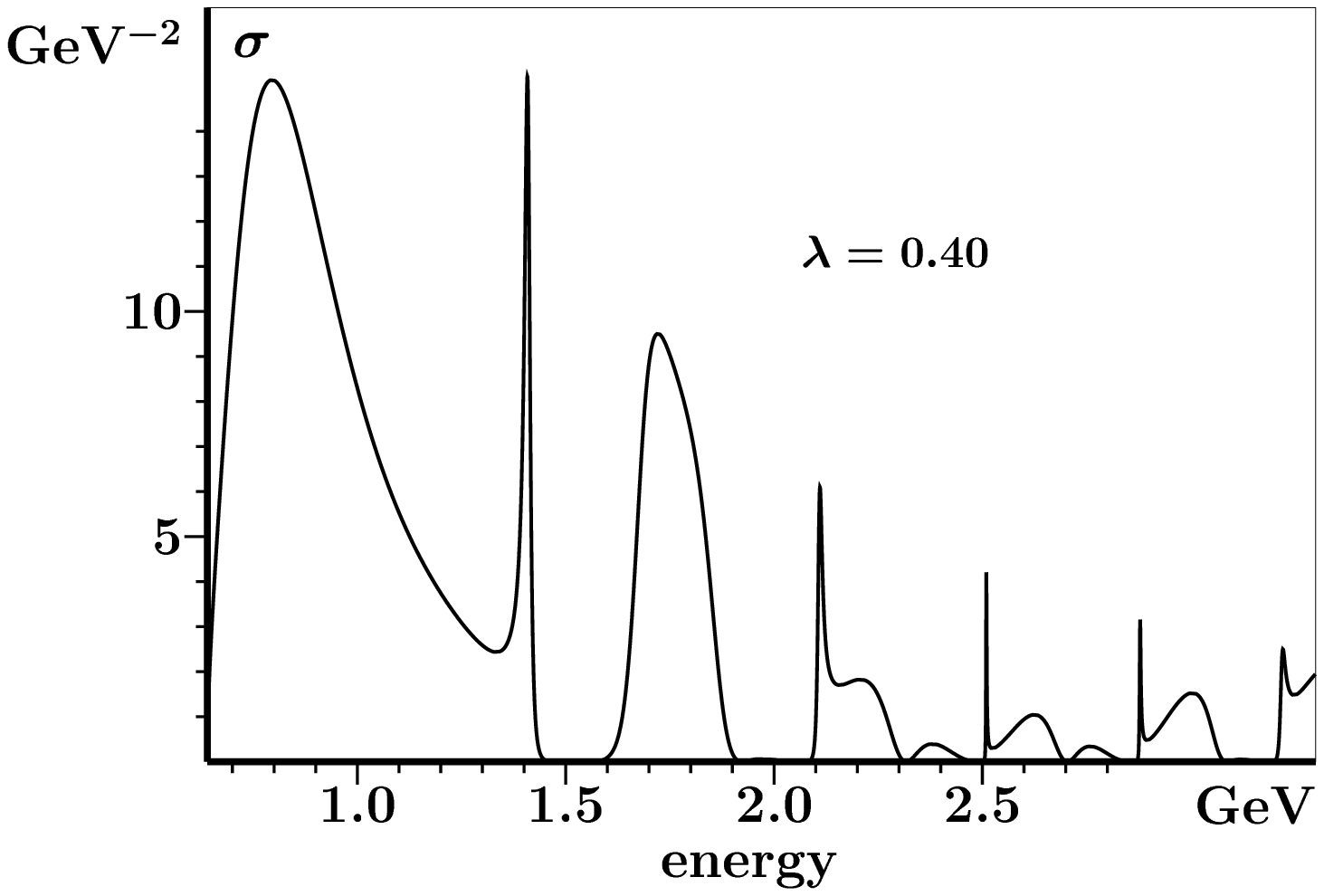}}}
\end{tabular}
\end{center}
\caption{The cross section of elastic $K\pi$ scattering in the harmonic
oscillator model for confinement, for various values of the coupling
$\lambda$.}
\label{crKpi}
\end{figure}

In Fig.~\ref{crKpi}, we display the cross section for harmonic-oscillator
confinement coupled to meson-meson scattering.
The parameters, i.e., the nonstrange and strange quark masses and the
oscillator frequency, are chosen (see Table~\ref{parameters}) such that the
\begin{table}[htbp]
\begin{center}
\caption{The relevant Nij\-me\-gen-model parameters from
Ref.~\protect\cite{PRD27p1527} (units are in GeV).
\label{parameters}}
\begin{tabular}{c|c|c|c|c}
$\omega$ & up/down mass ($m_{n}$) & $m_{s}$ & $m_{c}$ & $m_{b}$\\
\hline & & & & \\ [-0.2cm]
0.19 & 0.406 & 0.508 & 1.562 & 4.724
\end{tabular}
\end{center}
\end{table}
radial excitations come out at 1.39, 1.77, 2.15, 2.53, \ldots\ GeV.
In the (elastic) meson-meson scattering channel, we have chosen the
meson masses at 0.14 GeV (pion) and 0.5 GeV (Kaon).
The intensity of the coupling, parametrised by $\lambda$, is varied.
For $\lambda =0.1$ we are in the small-coupling limit (upper-left
picture of Fig.~\ref{crKpi}), for which a sum of BW amplitudes works
reasonably well. We also notice that, for $\lambda =0.1$, all peaks are near
the energy levels of the pure harmonic oscillator.
But as we increase the coupling, we observe that the cross section has
structures which are not of a BW type.
For $\lambda =0.4$, we see a dominant tructure at 0.8 GeV, which in a more
realistic approach \cite{ZPC30p615,EPJC22p493} can be identified with the
$K^{\ast}_{0}(800)$ resonance.
From the results of Fig.~\ref{crKpi}, we conclude that, for the
description of strong interactions, models are needed treating
the binding forces and the transition interactions on an equal footing.
Our proposal for elastic meson-meson scattering is to use the following
expression for the partial-wave phase shift:
\begin{equation}
\xrm{cotg}\left(\delta_{\ell}(p)\right)\; =\;
\fnd{\pi\lambda^{2}\mu p\;
\dissum{n=0}{\infty}
\fnd{{\cal J}^{\ast}_{n\ell}(p)\;{\cal N}_{n\ell}(p)}
{E(p)-E_{n\ell_{c}}}\; -\; 1}
{\pi\lambda^{2}\mu p\;
\dissum{n=0}{\infty}
\fnd{{\cal J}^{\ast}_{n\ell}(p)\;{\cal J}_{n\ell}(p)}
{E(p)-E_{n\ell_{c}}}}
\;\;\; .
\label{partialpshift}
\end{equation}
The details of this formula and its derivation can be found in
Refs.~\cite{AIPCP687p86,HEPPH0304105}. It expresses the fact that in strong
hadronic decay all possible confinement states contribute to the scattering
amplitude, through the factors ${\cal J}_{n\ell}(p)$ and
${\cal N}_{n\ell}(p)$, which are convolution integrals over all
possible confinement excitations, the transition potential, and
spherical Bessel resp.\ Neumann functions.
In the limit of small coupling $\lambda$, Eq.~(\ref{partialpshift})
reduces to a BW form, with a dominant contribution from the
nearest confinement state.

At this stage we may formulate our definition of resonances.
Structures like those depicted in Fig.~\ref{crKpi} are related
to complex-energy singularities in Eq.~(\ref{partialpshift}).
Those for which a pole in the complex-energy plane (in the nearest
Riemann sheet) can be identified are given by us the status of
resonance. For small coupling (see Fig.~\ref{crKpi} for $\lambda =0.1$),
Eq.~(\ref{partialpshift}) has poles near the confinement spectrum,
close to the real axis and with small negative imaginary parts.
However, poles can be found, too, with very large (up to several GeVs)
imaginary parts, though possibly having no measurable effect. The latter poles
can be attributed the status of \em mathematical \em \/resonance, but they may
be unobservable. Hence, whether or not we include them in our resonance list
depends on the accuracy of measurement, not on their actual existence.
For stronger coupling we see them showing up as genuine structures in the
$S$-wave cross section. Suddenly our spectrum has become richer than the
original confinement spectrum. This is the realm of strong interactions!

Expression (\ref{partialpshift}), which is the exact solution of
a simple dynamical model, is very compact and elegant. But the summations over
the radial excitations of the confined system do not converge fast. Hence, for
the numerical calculation of scattering quantities it is not suitable.
A more practical method has been described in Ref.~\cite{CPC27p377}.
Nevertheless, because of its close contact with the physical situation,
Eq.~(\ref{partialpshift}) is extremely transparent,
and useful as a starting point for making approximations.

For example, when we are studying a limited range of energies,
we may restrict the summation to only the few terms near by in energy,
characterised by $E_{n\ell_{c}}$ in Eq.~(\ref{partialpshift}),
and approximate the remaining terms by a simple function of $p$,
the relative meson-meson linear momentum.
Notice, however, that the resulting expression is only of the type of
a BW expansion if the strength of the transition, characterised by $\lambda$,
is small.
For larger values of $\lambda$, formula~(\ref{partialpshift}) behaves
very differently from BW expansions.

Another in practice very useful approximation is inspired by the radial
dependence of the transition potential, which was derived in
Ref.~\cite{ZPC21p291} for meson-meson scattering in the cases
$J^{P}=0^{-}$, $J^{P}=1^{-}$, and $J^{P}=0^{+}$.
We find there that it is peaked at short distances.
Moreover, if we also take into account that confinement wave functions
must be of short range, we may just define a transition radius $a$ and
approximate the spatial integrals in Eq.~(\ref{partialpshift}) by choosing a
spherical delta shell $\delta (r-a)$ for the transition potential.
Then we end up with the expression
\begin{equation}
\xrm{cotg}\left(\delta_{\ell}(p)\right)\;\approx\;
\fnd{2a^{4}\lambda^{2}\mu p\;
j_{\ell}(pa)\; n_{\ell}(pa)\;
\dissum{n=0}{\infty}
\fnd{\abs{{\cal F}_{n\ell_{c}}(a)}^{2}}
{E(p)-E_{n\ell_{c}}}\; -\; 1}
{2a^{4}\lambda^{2}\mu p\;
j^{2}_{\ell}(pa)\;
\dissum{n=0}{\infty}
\fnd{\abs{{\cal F}_{n\ell_{c}}(a)}^{2}}
{E(p)-E_{n\ell_{c}}}}
\;\;\; .
\label{partialpshifta}
\end{equation}
This way we have pulled the $p$ \/dependence outside the infinite summation
over the radial confinement spectrum, which
makes it much easier to handle truncations, as now the rest term does not
depend on $p$ \/and may thus be chosen constant.
\clearpage

\section{Open charm}

Our recent detailed analyses of scalar mesons have actually been
performed in the approximation of formula~(\ref{partialpshifta}).
For $DK$ elastic $S$-wave scattering, we adopt the same expression
that has been used for $K\pi$ as described in Ref.~\cite{EPJC22p493}.
We only have to substitute the confinement masses of $d\bar{s}$
by those for $c\bar{s}$, thereby using the parameters of
Table~\ref{parameters}.
For the ground state of the $J^{P}=0^{+}$ $c\bar{s}$ confinement spectrum,
the mass comes out at 2.545 GeV.
The first radial excitation is then 380 MeV (the model's level spacing) higher
in mass.  The summation over the higher radial excitations in
Eq.~(\ref{partialpshifta}) we approximate by a constant, normalised to 1.
The relative couplings of the ground state and the first radial
excitation to the $DK$ channel are 1.0 and 0.2, respectively \cite{EPJC22p493}.
Thus we obtain
\begin{equation}
\sum_{n=0}^{\infty}\;
\fnd{\abs{{\cal F}_{n,1}(a)}^{2}}
{E(p)-E_{n,1}}\;
\longrightarrow\;
\fnd{1.0}{E(p)-E_{0}}\; +\; \fnd{0.2}{E(p)-E_{1}}\; -\; 1
\;\;\;\;\;\xrm{GeV}^{\: 2}
\;\;\; ,
\label{DKSwave}
\end{equation}
where $E_{0}=2.545$ and $E_{1}=2.925$.
For Eq.~(\ref{partialpshifta}) with the substitution~(\ref{DKSwave}),
and furthermore for the physical value of $\lambda$ (meaning the value
that fits $K\pi$ elastic $S$-wave scattering from threshold up to 1.6 GeV
\cite{EPJC22p493}), we show the resulting elastic $S$-wave $DK$ cross
section in Ref.~\cite{AIPCP687p86}.
The corresponding pole positions in the complex-energy plane
for the $D^{\ast}_{s0}$ ground state and first radial excitation
are shown in Ref.~\cite{PRL91p012003}.
These are found at 2.28 GeV and ($2.78-i\, 0.093$) GeV, respectively .
We find no structure in the cross section in the region around 2.55 GeV,
where typical quark models predict $J^{P}=0^{+}$ $c\bar{s}$ states
\cite{PRD43p1679,PRD64p114004}, nor poles in the scattering amplitude.
On the contrary, employing confinement \em plus \em \/unitarisation,
we find structures in the cross section some 250 MeV below and 300--400 MeV
above that energy, which explains the observation of the (likely)
$D^{\ast}_{s0}$ ground state below the $DK$ threshold, and predicts for the
first radial excitation a resonance somewhere between 2.8 and 3.0 GeV, with a
width of roughly 100--200 MeV. We may also predict the scattering length for
$DK$ elastic $S$-wave scattering.  Keeping in mind that the $J^{P}=0^{+}$
$c\bar{s}$ ground state comes out a little bit too low in mass with our value
of $\lambda$, we estimate the theoretical error by varying $\lambda$.
For a conservative choice, we obtain for the scattering length in
$DK$ $S$ wave $a^0_{DK}=5\pm 1$ GeV$^{-1}$.
\clearpage

\section{Open beauty}

For $J^{P}\!=\!0^{+}$ systems containing one $b$ quark, we may repeat the
above-described procedure.
We determine the ground-state masses of the uncoupled systems
from the model parameters of Table~\ref{parameters}.
This results in 5.605, 5.707, and 6.761 GeV for $\bar{b}u/d$, $b\bar{s}$, and
$\bar{b}c$, respectively. The first radial excitations lie 0.38 GeV higher.
The remaining parameters $a$ and $\lambda$ are kept the same as before,
but now scaled with the reduced constituent $q\bar{q}$ mass $\mu_{qq}$, in
order to guarantee flavour invariance of the strong interactions
\cite{PRD59p012002}, i.e.,
\begin{equation}
a_{xy}\,\sqrt{\mu_{xy}}\; =\;\xrm{constant}
\;\;\;\;\xrm{and}\;\;\;\;
\lambda_{xy}\,\sqrt{\mu_{xy}}\; =\;\xrm{constant}
\;\;\;\; ,
\label{flavorinvariance}
\end{equation}
where $x$ and $y$ represent the two flavours involved \cite{HEPPH0310320}.
The constants in Eq.~(\ref{flavorinvariance}) are fixed by
\cite{HEPPH0110156} $a_{us}=3.2$ GeV$^{-1}$ and $\lambda_{us}=0.75$
GeV$^{-3/2}$ for $S$-wave $K\pi$ scattering.
Henceforth, we shall quote for the parameters $\lambda$ and $a$
their values scaled down to elastic $K\pi$ scattering.
The actual values of $\lambda$ and $a$ can, in each case, be obtained through
Eq.~(\ref{flavorinvariance}) and Table~\ref{parameters}.
\begin{figure}[htbp]
\begin{center}
\begin{tabular}{c}
\centerline{
\scalebox{0.65}{\includegraphics{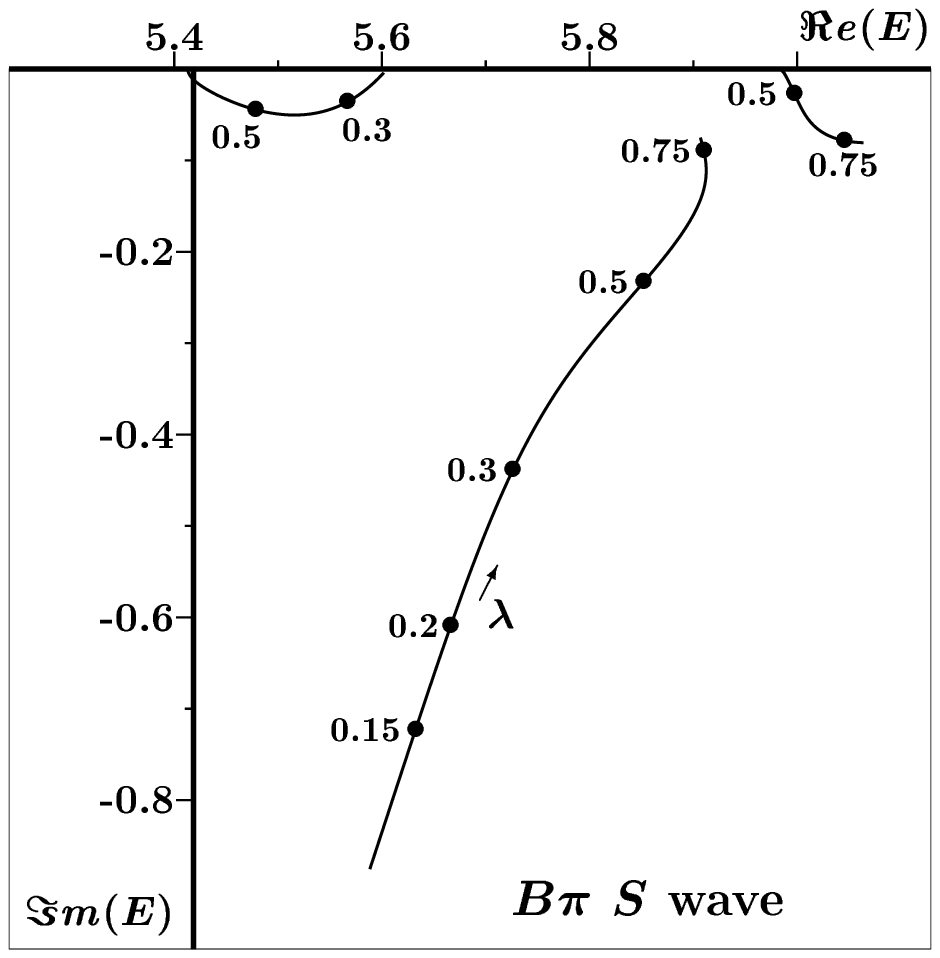}}
\scalebox{0.65}{\includegraphics{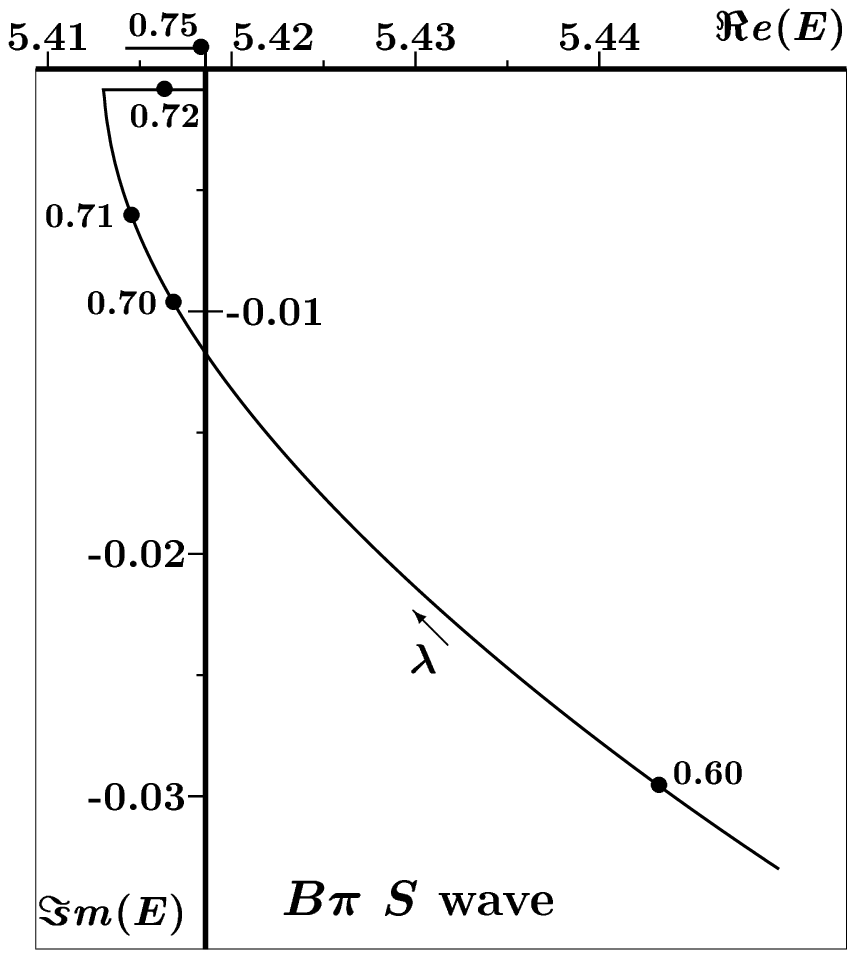}}}\\ [-8pt]
\hspace{37pt}{\bf (a)}\hspace{153pt} {\bf (b)}
\end{tabular}
\end{center}
\caption{a: Hypothetical pole movements for $J^{P}=0^{+}$ $B\pi$ elastic
scattering. We take $E_{0}=$ 5605 MeV and $E_{1}=$ 5985 MeV;
b: Details of the lower trajectory.}
\label{bn}
\end{figure}

In Fig.~\ref{bn} we show how in our model the scattering poles move
through the complex-energy plane for $B\pi$ elastic scattering,
when the coupling is varied.
For the lowest pole we obtain the very interesting result that the physical
pole (for $\lambda =0.75$) ends up on top of the threshold.
Within the accuracy of our model, this means that one has to expect
in experiment either a virtual, or a real bound state, just below
threshold.
Only through an accurate measurement of the scattering length
in the $B\pi$ $S$ wave, the precise position can be obtained experimentally.

For higher energies, we only find poles at energies far above 5.6 GeV.
This implies that we do not expect any resonance in the $S$ wave below roughly
5.9 GeV, where the confinement spectrum of $\bar{b}n$ has its first radial
excitation. In our model we obtain two resonances, each some 200 MeV wide,
and about 150 MeV apart.

\begin{figure}[htbp]
\begin{center}
\begin{tabular}{c}
\centerline{
\scalebox{0.65}{\includegraphics{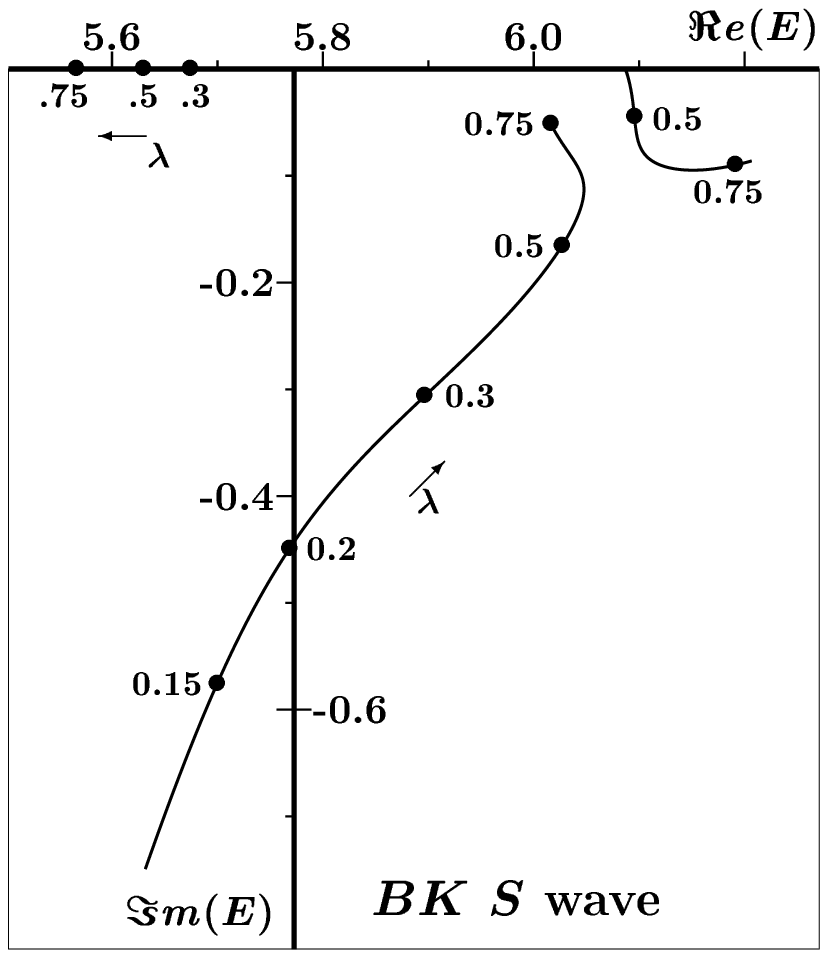}}
\scalebox{0.65}{\includegraphics{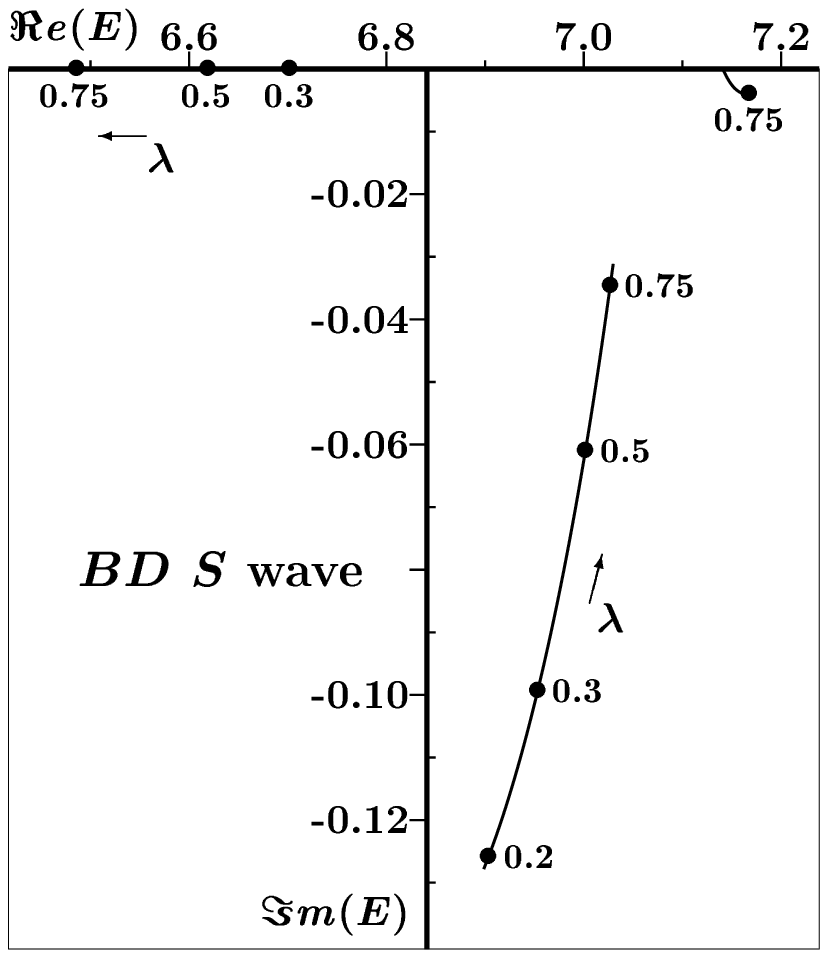}}}\\ [-8pt]
{\bf (a)}\hspace{160pt} {\bf (b)}
\end{tabular}
\end{center}
\caption{Hypothetical pole movements for $J^{P}=0^{+}$
$BK$ ({\bf a}) and $BD$ ({\bf b}) elastic scattering.
We take $E_{0}=$ 5707 MeV and $E_{1}=$ 6087 MeV for $b\bar{s}$ ({\bf a}), and
$E_{0}=$ 6761 MeV and $E_{1}=$ 7141 MeV for $\bar{b}c$ ({\bf b}).}
\label{bsbc}
\end{figure}

In Fig.~\ref{bsbc} we depict similar pole trajectories, but now for $BK$ and
$BD$ elastic scattering.
We find a $B_{s0}^{\ast}$(5570) bound state below the $BK$ threshold, which
experimentally should show up as a narrow $\bar{b}u/d$ resonance.
In the $BD$ case we find a $B_{c0}^{\ast}$(6490) bound state below threshold,
which represents a $b\bar{s}$ bound state. Also here, nothing is found at 5.707
or 6.761 GeV, where the confinement spectrum has its ground states.
Possible resonances above threshold and their widths can be read off from
Fig.~\ref{bsbc}.
We summarise these results in Table~\ref{bpoles}.
%\clearpage

\begin{table}[htbp]
\begin{center}
\caption{Bound states and resonances for $B\pi$, $BK$, and $BD$ elastic
$S$-wave scattering, as read from Figs.~\ref{bn} and \ref{bsbc}.
\label{bpoles}}
\begin{tabular}{c||c||c|c}
& ground state & \multicolumn{2}{c}{resonances}\\
$q\bar{q}$ & mass (GeV) & mass (GeV) & width (GeV)\\
\hline & & & \\ [-0.4cm]
$\bar{b}u/d$ & at $B\pi$ threshold & 5.90 & 0.2\\
& & 6.0 -- 6.1 & 0.2\\
\hline & & & \\ [-0.4cm]
$b\bar{s}$ & 5.57 & 6.0 -- 6.05 & 0.1\\
& & 6.2 & 0.2\\
\hline & & & \\ [-0.4cm]
$\bar{b}c$ & 6.49 & 7.0 -- 7.05 & 0.07\\
& & 7.18 & narrow
\end{tabular}
\end{center}
\end{table}
\clearpage

\section{Who is who?}

On applying the flavour-independence relations (\ref{flavorinvariance})
also to $S$-wave $DK$ scattering, we obtained an interesting result,
which we shall report here.

In our initial analysis \cite{PRL91p012003}, we identified the
$D^{\ast}_{sJ}(2317)^{+}$ meson with the lowest-lying extra pole,
which stems from the background, whereas the higher resonance
at about 2.9 GeV, with a width of some 150 MeV, was linked to the
confinement ground state.
However, if we choose for the delta-shell radius $a$ its corrected
flavour-independent value, then we find that the roles of these two poles
get interchanged.

\begin{figure}[htbp]
\centerline{\scalebox{0.95}{\includegraphics{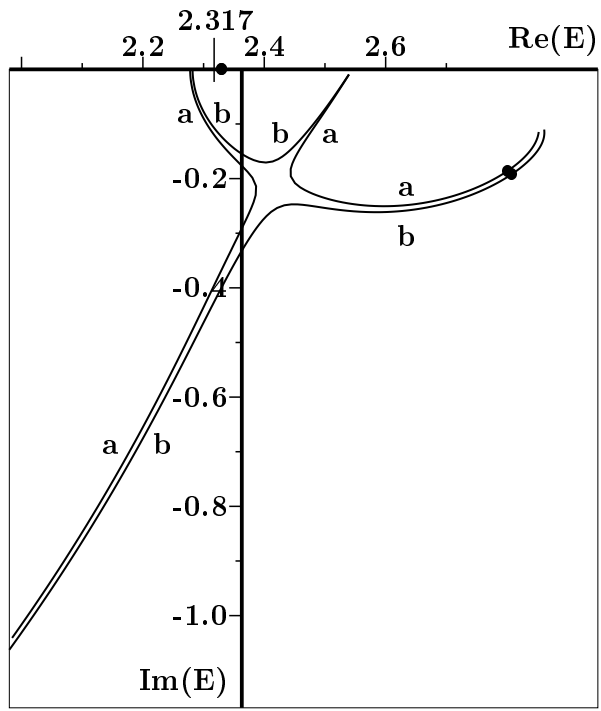}}}
\caption{Pole movements for $DK$ $S$-wave scattering, with
the flavour-invariant radius of the transition potential at the value
3.45 GeV$^{-1}$ before crossing (trajectories marked with {\bf a}),
and 3.40 GeV$^{-1}$ after crossing (trajectories marked with {\bf b}).
The $\bullet$s correspond to the {\it new} \/physical value of $\lambda$.}
\label{crossing}
\end{figure}

This phenomenon is depicted in Fig.~\ref{crossing},
where, upon a very small variation in $a$, the pole trajectories
interchange their physically relevant limiting points, which nevertheless
change only very little themselves.
As a consequence, we must conclude that the ``identity'' of a pole cannot
always be inferred with certainty. While in the small-coupling situation
(see e.g.\ the $\lambda\!=\!0.1$ case in Fig.~\ref{crKpi}) it is
unquestionable which poles are to be associated with the confinement spectrum,
for intermediate and large couplings one can always find other parameters in
Eq.~(\ref{partialpshift}) that, when varied simultaneously with the coupling,
lead to an identity change. Since intermediate-size couplings are typical for
strong hadronic decay, one must conclude that such processes are inherently
nonperturbative, and thus must be treated accordingly.

We should mention, moreover, that the flavour-independence
relation~(\ref{flavorinvariance}) mildly alters our previous results.
The $D^{\ast}_{sJ}(2317)^{+}$ pole now comes out at 2.32 GeV
(actually very close to the experimental mass), instead of at the published
\cite{PRL91p012003} value of 2.28 GeV. Furthermore, we find that the higher
resonance shifts even more considerably. Here we obtain a width of some
400 MeV instead of 150 MeV, and a central mass around 2.8 GeV, roughly
100 MeV below the initially proposed value.
\clearpage

\section{Summary and conclusions}

\begin{figure}[htbp]
\centerline{\scalebox{0.95}{\includegraphics{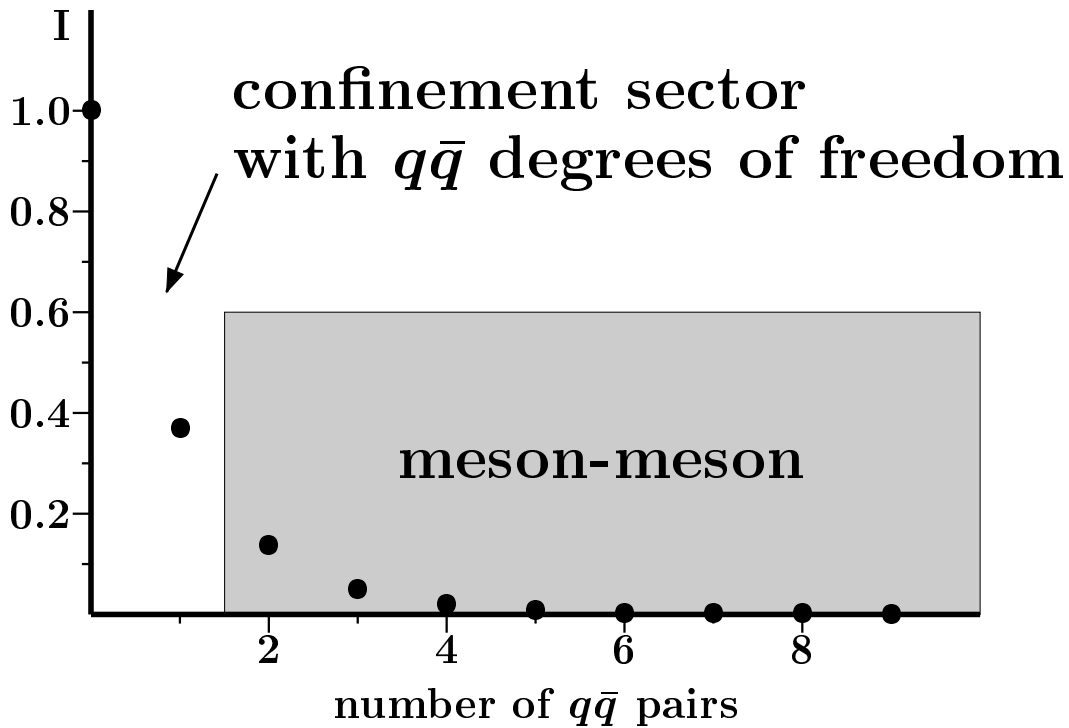}}}
\caption{
The weight \bm{I} \/of contributions to the dynamics of mesonic states,
for different configurations of \bm{q\bar{q}} pairs.}
\label{import}
\end{figure}

The main issue of this paper is schematically depicted in Fig.~\ref{import},
where we show how the various sectors of meson physics contribute to resonances
in meson-meson scattering.
Glue is probably dominant for the dynamics of meson-meson scattering.
It provides confinement, contributes dominantly to the effective masses of the
light quarks, and is also involved, though in a complicated and not entirely
understood fashion, in the creation of quark-antiquark pairs with vacuum
quantum numbers. The degrees of freedom of mesons are given by their
$q\bar{q}$ content. Higher-order configurations of $q\bar{q}$ \/pairs mediate
the transitions of a mesonic state to multi-meson states. Figure~\ref{import}
also indicates schematically the origin of deformation of the confinement
sector by the multi-meson sector, which results in mass shifts, resonance
widths, extra resonances, and bound states. For all that, processes
involving the creation of one or more $q\bar{q}$ pairs are generally strong,
and should \em not \em \/be handled perturbatively.

As to the spectroscopy of open-charm and open-beauty mesons, we predict several
bound states and resonances in $DK$, $B\pi$, $BK$, and $BD$ $\,S$~waves.
\clearpage

{\bf Acknowledgements}

We wish to thank the organisers of the Workshop,
in particular L\'{\i}dia Ferreira, for the warm hospitality.
We are also indebted to David Bugg and Frieder Kleefeld for very useful
discussions.
This work was partly supported by the {\it Funda\c{c}\~{a}o para a
Ci\^{e}ncia e a Tecnologia} \/of the {\it Minis\-t\'{e}rio da Ci\^{e}ncia
e do Ensino Superior} \/of Portugal, under contract numbers
CERN/\-FIS/\-43697/\-2001 and POCTI/\-FNU/\-49555/\-2002.

\end{document}